\documentclass[conference]{IEEEtran}
\IEEEoverridecommandlockouts

\usepackage{tabularx}
\usepackage{adjustbox}
\usepackage{multirow, makecell}
\usepackage{graphicx}
\usepackage{subcaption}
\usepackage{mwe}
\usepackage{enumitem}

\usepackage{hyperref}
\usepackage{academicons}

\usepackage{scalerel}
\usepackage{tikz}
\usetikzlibrary{svg.path}

\definecolor{orcidlogocol}{HTML}{A6CE39}
\tikzset{
  orcidlogo/.pic={
    \fill[orcidlogocol] svg{M256,128c0,70.7-57.3,128-128,128C57.3,256,0,198.7,0,128C0,57.3,57.3,0,128,0C198.7,0,256,57.3,256,128z};
    \fill[white] svg{M86.3,186.2H70.9V79.1h15.4v48.4V186.2z}
                 svg{M108.9,79.1h41.6c39.6,0,57,28.3,57,53.6c0,27.5-21.5,53.6-56.8,53.6h-41.8V79.1z M124.3,172.4h24.5c34.9,0,42.9-26.5,42.9-39.7c0-21.5-13.7-39.7-43.7-39.7h-23.7V172.4z}
                 svg{M88.7,56.8c0,5.5-4.5,10.1-10.1,10.1c-5.6,0-10.1-4.6-10.1-10.1c0-5.6,4.5-10.1,10.1-10.1C84.2,46.7,88.7,51.3,88.7,56.8z};
  }
}

\newcommand\orcidd[1]{\href{https://orcid.org/#1}{\mbox{\scalerel*{
\begin{tikzpicture}[yscale=-1,transform shape]
\pic{orcidlogo};
\end{tikzpicture}
}{|}}}}

\usepackage{cite}
\usepackage{amsmath,amssymb,amsfonts}
\usepackage{algorithmic}
\usepackage{graphicx}
\usepackage{textcomp}
\usepackage{xcolor}
\def\BibTeX{{\rm B\kern-.05em{\sc i\kern-.025em b}\kern-.08em
    T\kern-.1667em\lower.7ex\hbox{E}\kern-.125emX}}

\begin{document}

\title{MexPub: Deep Transfer Learning for \underline{Me}tadata E\underline{x}traction from German \underline{Pub}lications}

\author{

\IEEEauthorblockN{Zeyd Boukhers \quad Nada Beili \quad Timo Hartmann \quad Prantik Goswami \quad Muhammad Arslan Zafar}

\IEEEauthorblockA{\textit{Institute for Web Science and Technologies (WeST)} \\
\textit{University of Koblenz-Landau}\\
Koblenz, Germany \\
\{boukhers,nbeili,tihartmann,prantik,arslanzafar\}@uni-koblenz.de}

}

\maketitle

\begin{abstract}
Extracting metadata from scientific papers can be considered as a solved problem in NLP due to the high accuracy of state-of-the-art methods. However, this does not apply to German scientific publications, which have a variety of styles and layouts. In contrast to most of the English scientific publications that follow standard and simple layouts, the order, content, position and size of metadata in German publications vary greatly among publications. This variety makes traditional NLP methods fail to accurately extract metadata from these publications. In this paper, we present a method that extracts metadata from PDF documents with different layouts and styles by viewing the document as an image. We used Mask R-CNN that is trained on COCO dataset and finetuned with PubLayNet dataset that consists of ~200K PDF snapshots with five basic classes (e.g. text, figure, etc). We refine-tuned the model on our proposed synthetic dataset consisting of ~30K article snapshots to extract nine patterns (i.e. author, title, etc). Our synthetic dataset is generated using contents in both languages German and English and a finite set of challenging templates obtained from German publications. Our method achieved an average accuracy of around $90\%$ which validates its capability to accurately extract metadata from a variety of PDF documents with challenging templates.
\end{abstract}


\section{Introduction}
\label{sec:introduction}

With the advent of electronic publishing and digital libraries, the volume of available scientific literature increased significantly throughout the past decades~\cite{ivanyukovich_unsupervised_metadata_extraction}. Particularly, the Web of Science reported that as of August 2020, it comprises 171 million records of scientific publications. Associating these records with metadata allows understanding the data by the users, librarians, scholars and others. Therefore, making each article available with its metadata is important in all disciplines to ensure easy and fast access to the literature. For example, it allows digital libraries to provide their users with intelligent search engines and recommendation systems for related documents~\cite{tkaczyk_new_2017}. Furthermore, the availability and accessibility of academic metadata allow the development of semantic-enable services, such as authors’ profiling, bibliometrics, and scientific social network analysis. The most straightforward way to ensure the availability of this information is to collect them directly from the author(s) when submitting scientific documents. However, still, a significant part of bibliographic data in disciplines such as social science is not accessible via bibliographic databases and a vast amount of already existing scientific documents have incomplete or entirely missing metadata information~\cite{tkaczyk_new_2017}.

One way to overcome this problem is by a post hoc processing which is automatically extracting metadata from PDF documents. Consequently, several approaches for automatic metadata extraction from scientific documents~\cite{tkaczyk_new_2017,GROBID,lipinski_evaluation_2013,hui_han_automatic_2003,adefowoke_ojokoh_automated_2009}. have been proposed. These approaches are used by different digital libraries and publishers due to their stable accuracy when the layout and the structure of the PDF documents are standard. 
However, in a comparative evaluation of several metadata extraction tools including GROBID~\cite{GROBID}, Lipinski et al.~\cite{lipinski_evaluation_2013} found that while some tools achieve an accuracy of around 90\% on the title, they can extract abstract or date with accuracy only between $26\%$ and $75\%$. Moreover, state-of-the-art methods do not provide high accuracy when applied to non-English layouts, such as those of German scientific publications. Generally, English publications follow a standard and simple layout. In contrast, German scientific publications greatly vary in terms of the order, position and size of metadata. This is confirmed by the absence of a lot of German publications from bibliographic indices~\cite{colavizza2019citation,chi2014role}.

Over the past years, deep learning methods have dramatically improved the state-of-the-art in many domains, such as object recognition and object detection~\cite{lecun_deep_2015}. In line with Stahl et al.~\cite{stahl_deeppdf_2018}, we assume in this paper that scientific publications have an inherent structure that can be learnt by a deep learning network to distinguish among metadata patterns. Hence, this paper proposes a novel deep learning method for extracting metadata from scientific documents by treating PDF documents as images. The proposed method extracts information (e.g. author) by detecting its ROI in the image. To this end, we manually annotated a set of snapshots of each first page of 100 German scientific documents. Furthermore, we automatically generated a corpus consisting of around 44K annotated images of synthetic scientific papers based on the templates derived from the manual annotation. Using this dataset, we have fine-tuned an implementation of Mask R-CNN~\cite{dollar2017}  to distinguish between nine classes of metadata, namely; title, authors, journal, abstract, date, DOI, address, affiliation, and email addresses. The carried out evaluation demonstrates that our proposed model achieves a high average precision overall and across all classes. The main contributions of this paper are: I) Up to our knowledge, this is the first proposed model that tackles metadata extraction from PDF documents using a computer vision-based approach; II) Unlike conventional approaches, our model focuses on challenging and nonstandard layouts; III) We introduced a new and challenging synthetic dataset for metadata extraction; and IV) The experimental results on the proposed dataset demonstrate the effectiveness of \emph{MexPub}.

Following this section, Section ``\textit{Related Work}'' discusses the related works. The next section presents the proposed approach and Section ``\textit{Experiments}'' presents the conducted experiments and the obtained results that validate the effectiveness of our approach. Finally, Section ``\textit{Conclusion}'' concludes this paper and gives insight into future directions.
\section{Related Work}
\label{related}

Throughout the past decades, several research works have investigated the problem of extracting metadata from scientific literature~\cite{hui_han_automatic_2003,adefowoke_ojokoh_automated_2009,lipinski_evaluation_2013,yang_pipelines_2019,tkaczyk_new_2017,boukhers2019end}, where most of them tackle this problem on a text level.  Tkaczyk~\cite{tkaczyk_new_2017,tkaczyk2014cermine}  suggests that information extraction from scientific documents usually consists of multiple sub-tasks. First, the pre-processing stage which refers to parsing the input documents and preparing them for further analysis. The second step involves segmenting the content of a document into basic segments (e.g., text lines or blocks) and determining there order within the document. In the third step, the identified segments are classified into different types (e.g., paragraph, references or metadata). Lastly, parsing refers to the syntactic analysis of text strings that detects attributes of particular segments, such as references, affiliation strings, author full names, etc. The main problem of this multi-phase approach is aggregating errors over phases. Therefore, Boukhers et al~\cite{boukhers2019end} propose to combine these steps in a probabilistic approach that identifies reference strings directly from the raw text. 

One standard way to easily and accurately extract information from documents is by relying on text and layout rules~\cite{guo2011rule,stadermann2013extracting}. These approaches follow a set of predefined rules, which means that the layout of the paper has to be standard and known beforehand. Therefore, most of the earlier works addressed the problem of classifying segment strings in scientific documents using context-based classifiers such as Hidden Markov Models (HMMs)~\cite{takasu_bibliographic_2003} and Conditional Random Fields (CRF)~\cite{hosseini2019excite,GROBID} or models like Support Vector Machines (SVMs)~\cite{hui_han_automatic_2003,williams2014scholarly,bui2016pdf}. HMMs use a sequence of instances as input and learn a generative model to pair it with a sequence of states. Thereby, each state depends only on its immediate predecessor~\cite{tkaczyk_new_2017}. Peng and McCallum~\cite{peng_accurate_2004} point to the limitations of standard HMM models when dealing with multiple non-independent features. SVMs, on the other hand, can handle a large variety of independent features~\cite{tkaczyk_new_2017}, but lack the possibility of mapping a whole sequence of instances to a sequence of states~\cite{peng_accurate_2004}. To overcome this, Han et al.~\cite{hui_han_automatic_2003} employ a two-stage SVM based approach. They first classify each line of a scientific document’s header into one or more of 15 classes. Subsequently, an iterative convergence procedure uses the previously predicted class labels of a line’s neighbors to improve its classification.

As stated by Stahl et al.~\cite{stahl_deeppdf_2018}, more contemporary methods mostly utilize Conditional Random Fields (CRFs) to extract metadata from scientific papers. CRFs represent undirected graphical models trained to make inter-dependent predictions~\cite{peng_accurate_2004}. In this way, these models combine the advantages of both finite-state HMM and discriminative SVM techniques by allowing arbitrary, dependent features and joint inference over entire sequences. In their comparative evaluation of several metadata extraction methods, Lipinski et al.~\cite{lipinski_evaluation_2013} found that GROBID~\cite{GROBID}, a system using CRFs, performed significantly better than SVM-based approaches. Over recent years, deep learning-based methods~\cite{lecun_deep_2015} became the state-of-the-art in various machine learning tasks such as object recognition and natural language processing. Among them, several approaches address the problem of extracting information from pdf documents~\cite{liu_automatic_2018,prasad2018neural}, however, in the field of metadata extraction, to date, only a limited number of approaches include deep learning techniques~\cite{siegel_extracting_2018,stahl_deeppdf_2018,liu_automatic_2018,hao2016table}. Computer vision-based algorithms aim at recovering the properties of objects in images, such as shape, illumination, and color distributions~\cite{szeliski2010computer}. Siegel et al.~\cite{siegel_extracting_2018} used distantly supervised Neural Networks (NNs) to extract figures from scientific documents, while Stahl et al.~\cite{stahl_deeppdf_2018} propose a computer vision approach to recognise metadata patterns in images of scientific papers. They implemented a modified version of a Convolutional Neural Network (CNN) to classify content in scientific documents as one of two types, namely ”paragraphs” and ”non-paragraphs.” The authors use an implementation of U-Net, which allows assigning a label to each pixel of an image of a PDF document’s page. This CNN architecture is usually referred to as Semantic Segmentation and requires a relatively small amount of training data. Similarly, Liu et al.~\cite{liu_automatic_2018} implement a two-step deep learning approach using the image information of headers in scientific papers combined with the text content to classify the header’s components into one of eight types of metadata.
\section{Metadata Extraction from Publications (\emph{MexPub})}
\label{approach}

\subsection{Architecture}
Our model is an implementation of Mask R-CNN, as proposed by He et al.~\cite{dollar2017}. Mask R-CNN is an object instance segmentation model dedicated to detect objects within images on a pixel-by-pixel level. To this end, Mask R-CNN extends Faster RCNN, which has two outputs for each candidate object, a class label, and bounding box offset. This extension is made by adding a branch for predicting an object mask and using Region of Interest (RoI)-Align instead of RoI-Pooling. The binary object mask represents the position on a pixel-level of each object within its bounding box. Mask R-CNN is state-of-the-art segmentation approach that achieves faster training and inference than previous algorithms\cite{dollar2017}.\\ 

We built our model using Detectron2~\cite{wu2019detectron2}, a PyTorch-based object detection library. The library includes state-of-the-art implementations of MASK R-CNN models trained on the COCO dataset\footnote{\url{http://cocodataset.org}}. Specifically, we implemented Mask R-CNN with a ResNeXt \cite{Xie_aggregated2016}  back-bone architecture with Feature Pyramid Network (FPN) following~\cite{lin_fpn2016}. 



\begin{figure}[ht]
\centering
\includegraphics[width=7cm]{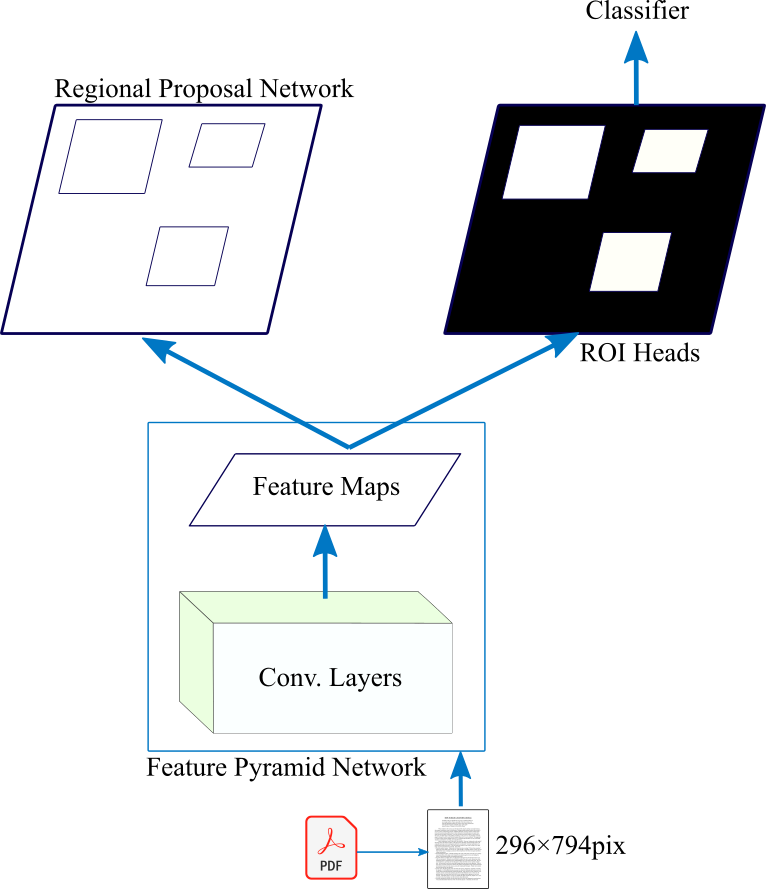}
\caption{Architecture of MASK R-CNN-FPN}
\label{fig:Mask_rcnn}
\end{figure}

As figure~\ref{fig:Mask_rcnn} illustrates, the model has three main blocks:  (i) an FPN,(ii) a Regional Proposal Network, and (iii) RoI Heads. As Table~\ref{tab:resnet} shows, The first block consists of a \textit{stem block} and four stages that contain multiple \textit{bottleneck blocks}. The \textit{stem block} down-samples the input image twice by 7x7 with stride 2 and max-pooling with stride 2, and extracts a feature map at scale 1/4. The remaining four stages contain \textit{bottleneck blocks}, each of which have three convolutional layers with kernel sizes 1x1, 3x3, and 1x1. These four stages consist of 3, 4, 23, and 3 \textit{bottleneck blocks}\footnote{\url{https://github.com/facebookresearch/detectron2/blob/e0bffda3f503bc4caa1ae2360520db3591fd291d/detectron2/modeling/backbone/resnet.py}} and extract feature maps at scales 1/4, 1/8, 1/16, and 1/32, respectively \cite{Xie_aggregated2016}. Finally, a max-pooling layer with kernel size 1 and stride 2 is added to the final stage of ResNeXt to output a feature map at scale 1/64 \cite{kaiming_maskrcnn2017}. The second block of the model consists of the Region Proposal Network (RPN), which proposes candidate object bounding boxes using the outputs of the five stages from the FPN as input. Finally, a fully convolutional mask prediction branch is added to the head  \cite{kaiming_maskrcnn2017}.
Finally, the RoI head uses fully-connected layers to generate fine-tuned box locations and classification results from multiple fixed-size features obtained by cropping and warping feature maps. The box head then filters out 100 boxes in maximum using non-maximum suppression (NMS).

\begin{table}[ht!]
\centering
\caption{ResNeXt-101 Architecture}

\label{tab:resnet}
 \begin{tabular}{c |c| c| c} 
 \hline
 Layer name & scale & kernel size & stride \\ [0.5ex] 
 \hline \hline
 stem & $1/4$ & $7\times7$ & $2$ \\ 
 \hline
 backbone $1$ & $1/4$ &  $\begin{bmatrix}
1\times1 \\
3\times3 \\
1\times1
\end{bmatrix} \times 3$ &  $1$ \\
 \hline
 backbone $2$ & $1/8$ &  $\begin{bmatrix}
1\times1 \\
3\times3 \\
1\times1
\end{bmatrix} \times 4$ & $1$ \\
 \hline
 backbone $3$ & $1/16$ &  $\begin{bmatrix}
1\times1 \\
3\times3 \\
1\times1
\end{bmatrix} \times 23$ & $1$ \\
 \hline
 backbone $4$ & $1/32$ &  $\begin{bmatrix}
1\times1 \\
3\times3\\
1\times1
\end{bmatrix} \times 3$ & $1$ \\  
 \hline
 max pooling layer&  $1/64$ &  
$1\times1$
 &  $2$ \\  
 \hline
\end{tabular}
\end{table}

\subsection{Transfer learning}
It is common practice in deep learning for computer vision to take advantage of convolutional networks pre-trained on large, training datasets and re-train them on a smaller, task-specific dataset to fine-tune weights and biases~\cite{mahajan_exploringlimits2018}. The process of transferring knowledge from one classification task to another is commonly referred to as transfer learning \cite{pan_transferlearning2010}.\\ \\

For our task, we adopt the training configurations of a source model\footnote{\url{https://github.com/hpanwar08/detectron2}} that builds on a Detectron2~\cite{wu2019detectron2} implementation of MASK R-CNN ResNeXt-101 32x8d FPN. The model was originally fine-tuned on a total of 191,832 images stemming from the PubLayNet dataset \cite{zhong2019publaynet}. PubLayNet consists of images of articles from PubMed Central\textsuperscript{TM} Open Access (PMCOA) and each image is annotated with regions of the following five classes: title, text, list, table, and figure. This model is considered to be suitable for extracting metadata from scientific papers since (i) its backbone was trained on the very large COCO dataset, (ii) it was fine-tuned on a large data set of scientific document snapshots,  which (iii) makes the task very similar to ours.

To adapt this model to the task of extracting metadata patterns from scientific documents, we, first,  modified the last layer of the source model to output the nine target classes (i.e. title, authors, journal, abstract, date, DOI, address, affiliation, and email addresses) instead of the original five classes. The empirical experiments on a subset of $10^{3}$ random samples from our training dataset demonstrates that the architecture with best performance is the one with 2 frozen layers and 15k iterations. Based on this finding, we refine-tuned the model using the full training dataset by setting the learning rate to $2.5\times10^{-3}$. The experiments were ran on Tesla P100 GPU using Google Colaboratory Pro\footnote{\url{https://colab.research.google.com}}.

\section{Experiments} \label{results}
\label{exp}

In this section, we discuss the conducted experiments and the used data. We also, compare \emph{MexPub} with a state-of-the-art approach and analyse the obtained results. Finally, we discuss the limitations of the approach and the potential solutions to overcome them.

\subsection{Data}

In this paper, we used a collection of 100 German scientific papers randomly selected from available publications in the SSOAR \footnote{\label{note1}\url{https://www.gesis.org/ssoar/home}} repository. The dataset includes various scientific layouts, which allowed us to train our model on a diverse data set and prevent the network from overfitting and memorizing layouts' exact details. In all the selected papers, most of the metadata is present on the first page. Therefore, we converted the first page of each paper (PDF) to JPEG. To improve our model's training procedure's computational speed, we resized all images to 596x794 pixels.

We manually annotated the regions corresponding to the desired metadata on a pixel-by-pixel level for each resized image. For this, we extended the metadata patterns proposed by Liu et al.~\cite{liu_automatic_2018}: title, author, affiliation, address, email, date, abstract, journal name, and DOI. We utilized the image annotation tool \textit{Labelme}\footnote{\url{http://labelme.csail.mit.edu/Release3.0/}} to draw polygons on the metadata sections and saved the annotations in a COCO\footnote{\url{http://cocodataset.org/home}} format. 

\begin{figure}[ht]
\centering

\includegraphics[width=6.5cm]{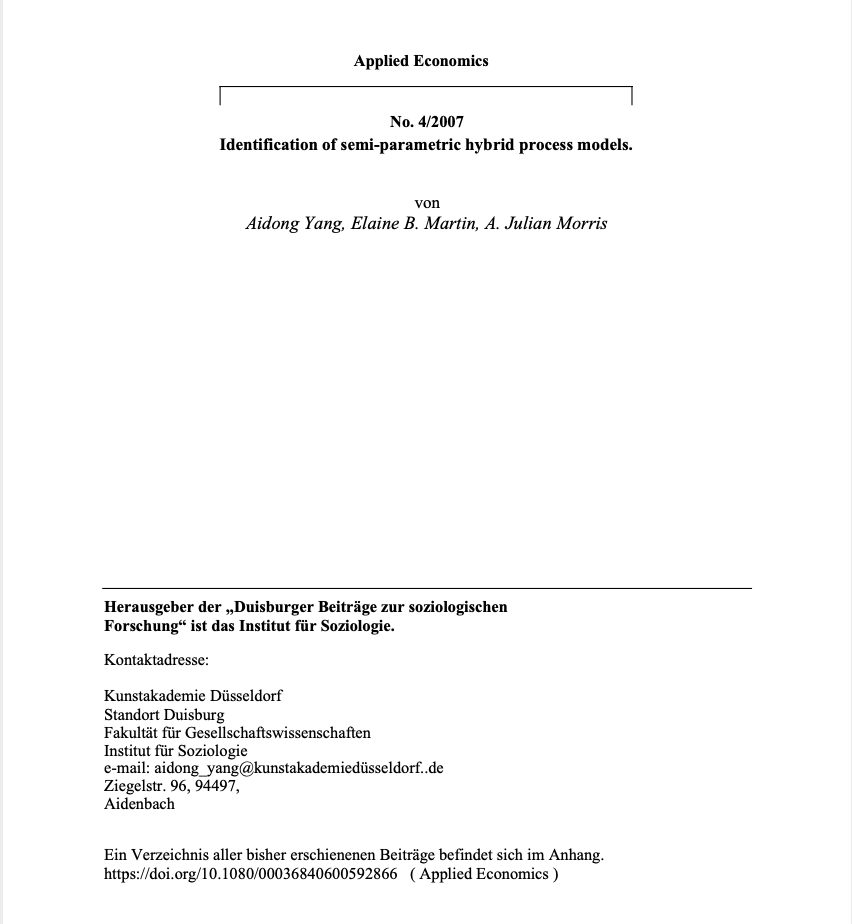}

\caption{Example image of the first page of a synthetic paper generated using metadata retrieved from the \texttt{DBLP,SSOAR} databases and scientific affiliations from \texttt{Wikipedia}.}
\label{fig:annotation}
\end{figure}

Due to the difficulty to annotate metadata on a pixel-by-pixel level, we extended our dataset by automatically generating synthetic papers based on the 28 most common layouts we identified during the mannual annotation phase. To this end, we randomly extracted metadata records from  \textit{SSOAR}\textsuperscript{\ref{note1}}, \textit{DBLP} \footnote{\url{https://dblp.org/xml/release/}}, and a list of scientific affiliations from Wikipedia\footnote{\url{https://de.wikipedia.org/wiki/Liste_der_Hochschulen_in_Deutschland}}. For each of these layouts, we generated an average of 1600 synthetic papers by randomly inserting metadata from the extracted metadata at their corresponding positions on the first page. This process ensures that samples are not just duplicated with different content but that relatively different templates are generated. For example, when longer titles are broke over two or several lines, the template slightly changes from the original one. This change increases under other factors such as multiple authors instead of one, different address style or when the date is absent.

For each of the synthetic sample, we again converted the first page to JPEG for each of the synthetic papers and automatically generated pixel-by-pixel annotations of the metadata patterns. Figure \ref{fig:annotation} shows an example of an automatically generated paper following the described process. Automatically generating synthetic papers allowed us to expand our dataset's size to around 44K papers with German and English content.

\begin{figure}[ht]
\centering

\includegraphics[scale=2]{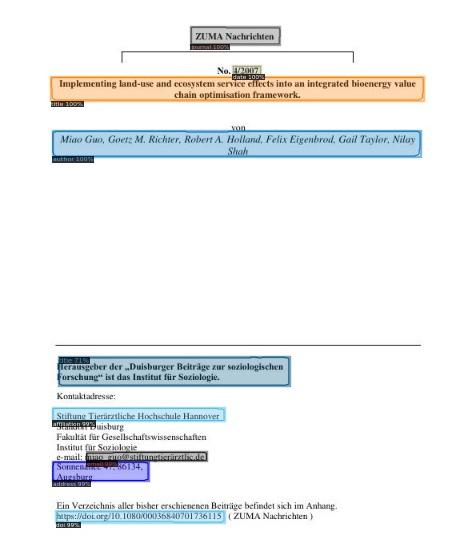}

\caption{Output example of our model for an image from the test dataset with all the present classes predicted correctly with very high confidence score except for the title class, for which the model predicted an additional false positive but with low confidence score.}
\label{fig:output}
\end{figure}

To evaluate the model, we randomly split our training data into 70\% training, 15\% validation, and 15\% test data. Figure~\ref{fig:output}  shows the model's output for an example image from the test set. The figure demonstrates the capability of the model to extract metadata from different layout styles, where it could perfectly extract metadata regions corresponding to all classes. A false positive title is extracted but with a low confidence score compared to the true positive classes. This means that a lot of extracted false-positive metadata can be discarded by introducing a confidence threshold. To quantitatively evaluate the model, Table \ref{tab:performance} summarizes the results for the validation and test sets after training the model with 15K iterations. \\

\vspace{-0.2cm}

%
\begin{table} [ht]
\centering
\caption{Performance of the model on validation and test data and trained with 15,000 iterations.}
\label{tab:performance}
\begin{tabular}{c |c| c} 
\hline
Average Precision         & Validation & Test \\
\hline \hline
\textbf{Overall}  & \textbf {90.363 }&\textbf {      90.167 }      \\ \hline
Abstract          &               97.363             &      97.567       \\\hline
Author            &                92.076          &       91.796      \\\hline
Email             &                87.035            &      85.938       \\\hline
Address           &               89.066            &       89.257      \\\hline
Date              &                 80.961          &    80.988         \\\hline
Journal           &               87.574          &      87.930       \\\hline
Affiliation       &                88.805           &     87.694        \\ \hline
DOI               &               95.027             &     94.573        \\ \hline
Title             &              95.360             &      95.761       \\ \hline
\end{tabular}
\end{table}

The model achieves an average precision (AP) for both the validation set ($AP~\thickapprox~90.363$) and the test set ($AP~\thickapprox~90.581$) when trained using 15,000 iterations. These AP-values validate our model's capability to accurately extract metadata from a diverse range of layouts and styles
of German scientific publications. The results reveal that the model performs exceptionally well for detecting abstracts, DOIs, and titles. We hypothesize that the high precision for DOIs might be due to its pattern having a persistent format across documents; It usually starts with ``\emph{https://doi,}'' followed by a chain of digits and punctuation. Regarding abstracts and titles, we assume that the high precision is partly explained by three factors; First, both patterns usually cover relatively large areas on the first pages of scientific documents, which the model better detects than smaller patterns. Specifically, our results show that for the validation set, the model achieves an $AP~\thickapprox~95.273$ for large pattern (areas $> 96$\textsuperscript{2} pixels), while for small pattern (areas $<$ 32\textsuperscript{2} pixels), the $AP~\thickapprox~74.173$. Second, in contrast to other metadata patterns, such as emails, addresses, or affiliation names, abstracts and titles are present in almost all training data samples. Finally, the positions of title and abstract are relatively consistent across layouts, whereas those of other metadata records vary more. Furthermore, we assume our model to benefit from the source model being trained to detect titles already. However, there are cases in which the model could not detect abstracts and titles accurately. First, for documents that do not contain an abstract, the model tends to falsely classify the first paragraph of the paper to be such. Second, the model tends to generate false positives for the title class when a document contains multiple areas of large-sized, bold text.
Moreover, the results reveal that dates are the most challenging pattern for the model to detect. Dates cover a relatively small area on the images, which we suppose to be one reason for the model's difficulties in detecting them. For both the validation and the test set, the model performs worse on objects with an area smaller than $32^2$ pixels than on larger ones. Furthermore, their positions on the page differ to a great extent across different layouts. Based on the assumption that the model factors in positional information, we assume that patterns with high positional variance across layouts are particularly challenging to recognize. 

\subsection{Comparison with the state-of-the-art}

Besides evaluating our model on the validation and test data sets, we compared it against the state-of-the-art method \textit{GROBID} \cite{GROBID}. To this end, we selected 100 scientific documents with different layouts from SSOAR\footnote{\url{https://www.gesis.org/ssoar/home}} that served as inputs to both pipelines. To allow for a fair comparison, we only included papers with a layout different from those we trained our model with. Note that these samples and their layouts are not seen in the training phase of our model. Moreover, since our model was trained only on German scientific papers' first pages, we also considered only the first pages from these samples. To evaluate the models' performances, we manually parsed all metadata records on the documents and used them as ground truth patterns. Since our model outputs the coordinates of bounding boxes for each detected metadata patterns, we extract the text from the PDF documents in the areas corresponding to these coordinates.

For both methods, we created confusion matrices with the predicted values as columns and the ground truth values as rows. To compute whether the extracted metadata matched the ground truth data, we used cosine similarity with a threshold of $0.85$. Whenever an extracted metadata record's cosine similarity was greater than or equal to the threshold when compared with the ground truth value, we consider it as true positive. The reasons of using soft matching by allowing a dissimilarity of $0.15$ is because some extracted patterns were more or less fine-grained than the ground truth. For example, the volume was not annotated as a part of the journal name, but both methods \emph{MexPub} and \textit{GROBID} include in in some cases while also eliminate it in other cases. Based on the confusion matrices, we computed average precision, average recall, and average F1 scores for the nine classes.\\ \\
 
Table~\ref{tab:Evaluation_metrics} depicts the results of the systems' evaluations. \emph{MexPub} is able to correctly predict titles and authors with an F1 score of $0.940$ and $0.750$, respectively. However, the results reveal that it produces a significant amount of false-positives for abstracts, which is explained by the low precision value for this class.

After analysing the obtained results, it is found that \emph{MexPub} tends to classify paragraphs that appear in the upper half of a document as abstracts, although they often represent introductions or other sections of the document's body. Similarly, the number of false-positives for journal and author is also relatively high. Furthermore, \emph{MexPub} did not predict any of the six present patterns corresponding to affiliations. Compared to \emph{MexPub}, \textit{GROBID} achieves a high precision, where it produces a significantly lower amount of false-positives across all classes present in the ground truth data. However, \textit{GROBID} fails at extracting a lot of true positive patterns which is reflected in the low recall for all classes. For some classes like ``address'' and ``journal'', \textit{GROBID} completely fails at extracting any of the corresponding patterns.

\begin{table}[h!]

\caption{Performances of GROBID and OUR MODEL regarding precision, recall, and F1-score per class. NaN values represent that the class was not present in the ground-truth data (e.g. DOI), or that a model is not designed to extract the corresponding pattern (e.g. date). For a fair comparison, the macro and micro averages are computed only using the classes associated with (\textsuperscript{*}) }
\centering
\label{tab:Evaluation_metrics}
\begin{adjustbox}{width=\columnwidth,center}
\begin{tabular}{@{}l|c |c| c|c|c|c}

\hline
\hline
            & \multicolumn{3}{c|}{\emph{MexPub}}                                                             & \multicolumn{3}{c}{GROBID}    \\ 
            \hline \hline
            & \multicolumn{1}{l|}{Precision} & \multicolumn{1}{l|}{Recall} & \multicolumn{1}{l|}{F1 Score} & Precision & Recall & F1 Score \\
            \hline
Title\textsuperscript{*}       & 0.934                         & \textbf{0.947}                      & \textbf{0.940}                        & \textbf{0.965}     & 0.577  & 0.723    \\ \hline
Author\textsuperscript{*}      & 0.670                         & \textbf{0.851}                      & 0.750                        & \textbf{0.982}     & 0.609  & \textbf{0.752}    \\ \hline
Journal\textsuperscript{*}     & \textbf{0.147}                         & \textbf{0.385}                      & \textbf{0.212}                        & 0.000     & 0.000  & 0.000      \\ \hline
Affiliation\textsuperscript{*} & 0.000                         & 0.000                      & 0.000                         & \textbf{1.000}     & \textbf{0.118}  & \textbf{0.210}    \\ \hline
Abstract\textsuperscript{*}    & 0.219                         & \textbf{0.833}                      & 0.346                        & \textbf{0.972}     & 0.593  & \textbf{0.739}    \\ \hline
DOI         & NaN                           & NaN                        & NaN                          & NaN       & NaN    & NaN      \\ \hline
Address\textsuperscript{*}     & \textbf{0.125}                         & \textbf{1.000}                      & \textbf{0.222}                        & 0.000       & 0.000  & 0.000      \\ \hline
Email\textsuperscript{*}       & 0.000                          & 0.000                        & 0.000                          & \textbf{1.000}     & \textbf{0.500}  & \textbf{0.667}    \\ \hline
Date        & 0.250                         & 1.000                      & 0.400                        & NaN       & NaN  & NaN      \\ \hline\hline

\textbf{Macro average}        & 0.299                         & \textbf{0.574}                     & 0.353                       & \textbf{0.703}       & 0.342  & \textbf{0.442}      \\ \hline

\textbf{Micro average}        & 0.558                         & \textbf{0.754}                     & \textbf{0.613}                       & \textbf{0.766}       & 0.447  & 0.559      \\ \hline

\end{tabular}
\end{adjustbox}
\end{table}

Although \emph{MexPub} does not consider any contextual features, it could achieve a good results on unseen documents with completely different layouts. However, the main limitation of this model is its low generalizibility to publications with a significantly different structure. Therefore, we assume that by integrating contextual/textual features, the model can perform better.

\section{Conclusion}
\label{conclusion}
In this paper, we proposed \emph{MexPub} that automatically extracts metadata from scientific papers using deep learning. Contrary to conventional approaches, \emph{MexPub} treats the PDF document as image and extracts metadata on a pixel-by-pixel level. For this, we adopted a deep learning model dedicated for object detection and retrained to detect patterns such as \textit{text} and \textit{figures} from PDF documents. We refine-tuned this model using the synthetic dataset proposed in this paper in order to extract fine grained patterns (e.g. title, author). The experimental results validates the capability of this approach to extract metadata from different layouts. 

For future work, we will train the model on a greater variety of layouts  to improve its generalizability. Using a larger amount of training data can further optimize the model's performance, especially concerning small-sized patterns, which pose a particular challenge to the model. Moreover, in some cases, \emph{MexPub} outputs multiple patterns of the same class for a given image when it should not, which leads to false positives. Future research will incorporate text-based processing in a joint neural network architecture. The visual part of this model is supposed to capture the structural characteristics of the PDF document, while the textual part is supposed to capture the semantic and contextual characteristics.

\bibliographystyle{plain}
\bibliography{acmart.bib}

\end{document}